\documentclass{article}
\usepackage{amsfonts}
\usepackage{amssymb}
\usepackage{amsmath}
\usepackage{amsthm}

\begin{document}

\title{\bf Dealing with ghost-free massive gravity\\ without explicit square roots of matrices}

\author{Alexey Golovnev,  Fedor Smirnov\\
{\it Faculty of Physics, St. Petersburg State University,}\\ 
{\it Ulyanovskaya ul., d. 1, Saint Petersburg 198504, Russia,}\\
{\small agolovnev@yandex.ru} \qquad {\small  sigmar40k@yandex.ru}}
\date{}

\maketitle

\begin{abstract}

In this paper we entertain a simple idea that the action of ghost free massive gravity (in metric formulation) depends not on the full structure of the square root of a matrix but rather on its invariants given by the elementary symmetric polynomials of the eigenvalues. In particular, we show how one can construct the quadratic action around Minkowski spacetime without ever taking the square root of the perturbed matrix. The method is however absolutely generic. And it also contains the full information on possible non-standard square roots coming from intrinsic non-uniqueness of the procedure. In passing, we mention some hard problems of those apocryphal square roots in the standard approach which might be better tackled with our method. The details of the latter are however deferred to a separate paper.

\end{abstract}
\section{Introduction}

The theory of General Relativity enjoys a superb agreement with experimental data all over a wide variety of scales. However, in the realm of cosmology we have a number of uneasy points including the origin of Dark Energy and the nature of Dark Matter. It gave rise to a plenitude of attempts to formulate a viable infrared modification of gravity which would hopefully do better in cosmology than GR. In particular, one of such directions which recently became very popular hinges upon giving a mass to the graviton.

The early days of massive gravity witnessed an almost detective story which starts from the original paper by Fierz and Pauli $\cite{FP}$ which presented the linearised ghost-free massive deformation around flat space, and goes through infamous vDVZ discontinuity \cite{MMYM,LGT} of its massless limit, to the potential resolution via Vainshtein mechanism \cite{VM, IVM}, and almost simultaneously to the claim of unavoidable reappearance of the ghost at non-linear level \cite{Deser}, and finally to the ultimate proposal by de Rham, Gabadadze and Tolley \cite{GFPA,RMG,Theo,NLAMG,RGPMG}. The model requires an additional (fiducial) metric which can either be Minkowski $\eta_{\mu\nu}$ as in the first papers on the subject, or can be arbitrary \cite{GFMGGRM,CSCAG} and even dynamical with its own Einstein-Hilbert term \cite{BGGMG} thereby producing a full-fledged bimetric gravity. 

An ugly feature of the model is that the interaction potential is made of $\sqrt{g^{-1}f}$, the square root of the matrix $g^{\mu\alpha}f_{\alpha\nu}$ which, strictly speaking, lacks both guaranteed existence (in the class of real matrices) and uniqueness, see also \cite{me,Marco}.  In this paper we present a method of dealing with massive gravity without explicitly taking the square root of the matrix. In Section 2 we describe the action of massive gravity and its second order expansion around flat space. In Section 3 we introduce the formalism of elementary symmetric polynomials of the eigenvalues, and also explain the problems with non-standard square roots in the usual formulation. In Section 4 we apply our method to quadratic gravity around flat space. Finally, in Section 5 we conclude.

\section{Massive gravity}

We consider the action of massive gravity in the following form: 

\begin{equation}
\label{TA}
S=\int d^N x \sqrt{-g} \left(R
+m^2 \sum_{n=0}^N \beta_n e_n(\sqrt{g^{-1}\eta})\right)
\end{equation}
where the spacetime is $N$-dimensional with metric $g_{\mu\nu}$, $R$ is its scalar curvature, and $e_n(\mathcal M)$'s are elementary symmetric polynomials of the eigenvalues $\lambda_i$ of the matrix $\mathcal M^{\mu}_{\nu}$:
\begin{equation}
e_n\equiv\sum\limits_{i_1<i_2<\ldots<i_n}\lambda_{i_1}\lambda_{i_2}\cdots\lambda_{i_n}
\end{equation}
and $e_0\equiv 1$ by definition. We see that $\beta_0$ gives a pure contribution to the cosmological constant, while $e_N(\sqrt{g^{-1}\eta})=\frac{1}{\sqrt{-g}}$ adds a mere constant to the action, and therefore it is totally irrelevant unless one wants to have a dynamical metric $f_{\mu\nu}$ instead of $\eta$ for which it would contribute to its own cosmological constant. Terms with $\beta_1, \ldots, \beta_{N-1}$ make up the potential term for the graviton.

Obviously, these polynomials can be described as coefficients in the characteristic polynomial of $\mathcal M$:
\begin{equation}
{\rm det}\left({\mathcal M}-\lambda{\mathbb I}\right)=\prod\limits_{n=1}^N\left(\lambda_i-\lambda\right)=\sum_{n=0}^N (-\lambda)^{N-n}\cdot e_n(\mathcal M).
\end{equation}
In particular, $e_1$ is the ordinary trace
\begin{equation}
\label{1L}
e_1(\mathcal M)=\sum_i\lambda_i=[\mathcal M]
\end{equation}
where $[\mathcal M]$ stands for the trace of $\mathcal M$. In other words, we have a shorthand notation which reads $[\mathcal M]\equiv {\mathcal M}^{\mu}_{\mu}$,  $[\mathcal M]^2\equiv ({\mathcal M}^{\mu}_{\mu})^2$,  $[\mathcal M^2]\equiv {\mathcal M}^{\mu}_{\nu}{\mathcal M}^{\nu}_{\mu}$, etc. Then we have
\begin{equation}
\label{2L}
e_2(\mathcal M)=\sum_{i<j}\lambda_i \lambda_j=\frac{1}{2}\left(\left(\sum_i\lambda_i\right)^2-\sum_i\lambda_i^2\right)=\frac{1}{2}\left([\mathcal M]^2-[\mathcal M^2]\right).
\end{equation}
And one can prove a simple recurrent relation
\begin{equation}
\label{recurrent}
e_n(\mathcal M)=\frac{1}{n}\sum\limits_{i=1}^n (-1)^{i-1}[\mathcal M^i]\cdot e_{n-i}(\mathcal M).
\end{equation}
with $e_n=0$ for $n>N$.

We will be interested only in the case of $N=4$, for which we get from (\ref{recurrent})
\begin{equation}
\label{3L}
e_3(\mathcal M)=\frac{1}{6}\left([\mathcal M]^3-3[\mathcal M][\mathcal M^2]+2[\mathcal M^3]\right)
\end{equation}
and also
\begin{equation}
\label{4L}
e_4(\mathcal M)= \frac{1}{24}\left([\mathcal M]^4-6[\mathcal M]^2[\mathcal M^2]+3[\mathcal M^2]^2+8[\mathcal M][\mathcal M^3]-6[\mathcal M^4]\right)=\text{det}(\mathcal M).
\end{equation}
The relevant parameters are $\beta_1$, $\beta_2$, and $\beta_3$. The mass parameter $m$ corresponds to the mass scale of the graviton if the largest of $\beta_i$'s (for $i=1,2,3$) is of order one.

In this paper we would be interested in linearised gravity around Minkowski spacetime, so that we take
$g_{\mu\nu}=\eta_{\mu\nu}+h_{\mu\nu}$
with a small perturbation $h$ to the metric. We will raise and lower the indices of $h$ by $\eta$. And then $h^{\mu\nu}$ gives the linear variation of $g^{-1}$ with inversed sign
$g^{\mu\nu}=\eta^{\mu\nu}-h^{\mu\nu}+{\mathcal O}(h^2)$,
or with a better accuracy we have
\begin{equation}
g^{\mu\alpha}\eta_{\alpha\nu}=\delta^{\mu}_{\nu}-h^{\mu}_{\nu}+h^{\mu\alpha}h_{\alpha\nu}+{\mathcal O}(h^3).
\end{equation}

In the standard approach, the square root matrix $\sqrt{g^{-1}\eta}$ would be found explicitly assuming the trivial root of the unity matrix: $\sqrt{\mathbb I}=\mathbb I$. Then the first terms of the Taylor expansion
$$\sqrt{{\mathbb I} - H}={\mathbb I} -\frac12 H - \frac18 H^2 + {\mathcal O}(H^3)$$ 
with $H=h-h^2+{\mathcal O}(h^3)$ give the desired result when substituted into (\ref{1L}), (\ref{2L}), (\ref{3L}), and (\ref{4L}):
\begin{eqnarray}
\label{e1}
e_1(\sqrt{g^{-1}\eta}) & = & 4-\frac12 h^{\mu}_{\mu}+\frac38 h_{\mu\nu}h^{\mu\nu}+{\mathcal O}(h^3),\\
\label{e2}
e_2(\sqrt{g^{-1}\eta}) & = & 6-\frac32 h^{\mu}_{\mu}+\frac18 (h^{\mu}_{\mu})^2+h_{\mu\nu}h^{\mu\nu}+{\mathcal O}(h^3),\\
\label{e3}
e_3(\sqrt{g^{-1}\eta}) & = & 4-\frac32 h^{\mu}_{\mu}+\frac14 (h^{\mu}_{\mu})^2+\frac78 h_{\mu\nu}h^{\mu\nu}+{\mathcal O}(h^3),\\
\label{e4}
e_4(\sqrt{g^{-1}\eta}) & = & 1-\frac12 h^{\mu}_{\mu}+\frac18 (h^{\mu}_{\mu})^2+\frac14 h_{\mu\nu}h^{\mu\nu}+{\mathcal O}(h^3).
\end{eqnarray}

Of course, the last expression (\ref{e4}) can also be derived from
$e_4(\sqrt{g^{-1}\eta})=\frac{1}{\sqrt{-g}}$ where
\begin{equation}
\label{-g}
\sqrt{-g}=1+\frac12 h^{\mu}_{\mu}+\frac18 (h^{\mu}_{\mu})^2-\frac14 h_{\mu\nu}h^{\mu\nu}+{\mathcal O}(h^3).
\end{equation}

Quadratic approximations to the $\beta_i$ terms in the action (\ref{TA}) are easily given by multiplying (\ref{e1}) -- (\ref{e3}) by (\ref{-g}):
\begin{eqnarray}
\sqrt{-g}\cdot e_1(\sqrt{g^{-1}\eta}) & = & 4+\frac32 h^{\mu}_{\mu}+\frac14 (h^{\mu}_{\mu})^2-\frac58 h_{\mu\nu}h^{\mu\nu}+{\mathcal O}(h^3),\\
\sqrt{-g}\cdot e_2(\sqrt{g^{-1}\eta}) & = & 6+\frac32 h^{\mu}_{\mu}+\frac18 (h^{\mu}_{\mu})^2-\frac12 h_{\mu\nu}h^{\mu\nu}+{\mathcal O}(h^3),\\
\sqrt{-g}\cdot e_3(\sqrt{g^{-1}\eta}) & = & 4+\frac12 h^{\mu}_{\mu}-\frac18 h_{\mu\nu}h^{\mu\nu}+{\mathcal O}(h^3),
\end{eqnarray}
$\sqrt{-g}\cdot e_4(\sqrt{g^{-1}\eta})=1$ exactly, and of course $\sqrt{-g}\cdot e_0=\sqrt{-g}$ given by (\ref{-g}).

In this form, the Fierz-Pauli structure of the potential term is not yet obvious. However, we see that there is a non-vanishing first order contribution to the action around Minkowski:
$$V(h)\equiv m^2 \sum_{n=0}^N \sqrt{-g}\cdot\beta_n e_n(\sqrt{g^{-1}\eta})=V(0)+m^2\left(\frac12\beta_0+\frac32\beta_1+\frac32\beta_2+\frac12\beta_3\right)h^{\mu}_{\mu}+{\mathcal O}(h^2)$$
In order for the flat space to be a solution, we require it vanish which gives a condition
$$\beta_0=-3\beta_1-3\beta_2-\beta_3.$$
Being plugged back into the action, it yields the familiar result:
$$V(h)-V(0)=\frac{m^2}{8}\left(\beta_1+2\beta_2+\beta_3\right)\cdot\left(h^{\mu\nu}h_{\mu\nu}-(h^{\mu}_{\mu})^2\right)+{\mathcal O}(h^3).$$

Note that we followed the usual path. However, these calculations can be simplified by employing the well-known symmetry of bimetric theory $g_{\mu\nu}\leftrightarrow f_{\mu\nu}$, $\beta_n\leftrightarrow\beta_{N-n}$. It comes from the fact that $e_n(\mathcal M^{-1})$ is a polynomial of $\frac{1}{\lambda_i}$ which can be obtained from $e_{N-n}(\mathcal M)$ by dividing over ${\rm det}\mathcal M$. In particular,
$$\sqrt{-g}\cdot e_3(\sqrt{g^{-1}\eta})=e_1(\sqrt{\eta^{-1}g})=e_1(\sqrt{{\mathbb I}+h})=4+\frac12 [h]-\frac18 [h^2]+{\mathcal O}(h^3)$$
which also explains the mysterious disappearance of the $(h^{\mu}_{\mu})^2$-term from $\sqrt{-g}\cdot e_3(\sqrt{g^{-1}\eta})$.

\section{Relating $e_n(\sqrt{g^{-1}\eta})$ to $e_n(g^{-1}\eta)$}

Now we want to find $e_n(\sqrt{g^{-1}\eta})$'s without calculating the matrix explicitly. The main observation for that is the following:
\begin{multline}
\label{relation}
\sum_{n=0}^N (-\lambda^2)^{N-n}\cdot e_n(\mathcal M^2)={\rm det}\left({\mathcal M^2}-\lambda^2{\mathbb I}\right)={\rm det}\left(({\mathcal M}-\lambda{\mathbb I})\cdot({\mathcal M}+\lambda{\mathbb I})\right)\\
={\rm det}\left({\mathcal M}-\lambda{\mathbb I}\right)\cdot{\rm det}\left({\mathcal M}+\lambda{\mathbb I}\right)=\left(\sum_{k=0}^N (-\lambda)^{N-k}\cdot e_k(\mathcal M)\right)\cdot\left(\sum_{m=0}^N \lambda^{N-m}\cdot e_m(\mathcal M)\right).
\end{multline}
Comparing the powers of $\lambda$ on the opposite sides, we see a trivially satisfied relation
$$\sum\limits_{k+m=2n+1}(-1)^k e_k(\mathcal M) e_m(\mathcal M)=0$$
and also deduce a very important equality:
\begin{equation}
\sum\limits_{k+m=2n}(-1)^k e_k(\mathcal M) e_m(\mathcal M)=(-1)^n e_n(\mathcal M^2)
\end{equation}
which relates the elementary symmetic polynomials of an arbitrary matrix $\mathcal M$ and its square $\mathcal M^2$.
In particular, in the $4$-dimensional case we have
\begin{eqnarray}
\label{eqsta}
e_1(\mathcal M^2) & = & e_1^2(\mathcal M)-2e_2(\mathcal M),\\
e_2(\mathcal M^2) & = & e_2^2(\mathcal M)-2e_1(\mathcal M)e_3(\mathcal M)+2e_4(\mathcal M),\\
e_3(\mathcal M^2) & = & e_3^2(\mathcal M)-2e_2(\mathcal M)e_4(\mathcal M),\\
\label{eqend}
e_4(\mathcal M^2) & = & e_4^2(\mathcal M).
\end{eqnarray}

Our final aim is ${\mathcal M^2}=g^{-1}\eta$. However, let us first consider the simplest example of ${\mathcal M}^2=\mathbb I$. Equations (\ref{eqsta}) -- (\ref{eqend}) take the form of
\begin{eqnarray*}
4 & = & e_1^2(\sqrt{\mathbb I})-2e_2(\sqrt{\mathbb I}),\\
6 & = & e_2^2(\sqrt{\mathbb I})-2e_1(\sqrt{\mathbb I})e_3(\sqrt{\mathbb I})+2e_4(\sqrt{\mathbb I}),\\
4 & = & e_3^2(\sqrt{\mathbb I})-2e_2(\sqrt{\mathbb I})e_4(\sqrt{\mathbb I}),\\
1 & = & e_4^2(\sqrt{\mathbb I}).
\end{eqnarray*}
They are fairly simple to analyse and admit a number of solutions which are listed below.

First solution is the most obvious one $e_1(\sqrt{\mathbb I})=\pm 4$, $e_2(\sqrt{\mathbb I})=6$, $e_3(\sqrt{\mathbb I})=\pm 4$, $e_4(\sqrt{\mathbb I})=1$  which corresponds to the trivial square root
\begin{equation*}
\sqrt{\mathbb I}=\pm\left(
\begin{matrix}
1 & 0 & 0 & 0 \\
0 & 1 & 0 & 0 \\
0 & 0 & 1 & 0\\
0 & 0 & 0 & 1
\end{matrix}\right).
\end{equation*}
It is what everybody is used to. However, it's not the end of the story.

Second solution reads $e_1(\sqrt{\mathbb I})=0$, $e_2(\sqrt{\mathbb I})=-2$, $e_3(\sqrt{\mathbb I})=0$, $e_4(\sqrt{\mathbb I})=1$ and encodes another square root 
\begin{equation*}
\sqrt{\mathbb I}=\pm\left(
\begin{matrix}
1 & 0 & 0 & 0 \\
0 & 1 & 0 & 0 \\
0 & 0 & -1 & 0\\
0 & 0 & 0 & -1
\end{matrix}\right)
\end{equation*}
together with all its similarity transformations since
$\left(\mathcal C\cdot \sqrt{\mathbb I}\cdot\mathcal C^{-1}\right)^2=\mathcal C\cdot(\sqrt{\mathbb I})^2\cdot\mathcal C^{-1}=\mathbb I$ for any non-degenerate matrix $\mathcal C$.

Finally, third solution with opposite sign of determinant $e_1(\sqrt{\mathbb I})=\mp 2$, $e_2(\sqrt{\mathbb I})=0$, $e_3(\sqrt{\mathbb I})=\pm 2$, $e_4(\sqrt{\mathbb I})=-1$ is possible. It features yet another matrix
\begin{equation*}
\sqrt{\mathbb I}=\pm\left(
\begin{matrix}
1 & 0 & 0 & 0 \\
0 & -1 & 0 & 0 \\
0 & 0 & -1 & 0\\
0 & 0 & 0 & -1
\end{matrix}\right),
\end{equation*}
again together with all its similarity transformations.

In what follows we will use the simplest (and the most important) choice of the first solution. We are about to show that our approach can easily reproduce the behaviour of the usual massive gravity model. The non-standard square roots will be discussed elsewhere \cite{paper2}. However, it is important to mention that in the standard language the perturbations around those choices all critically ill-defined. Indeed, one easily checks that for a block-diagonal matrix
$\left(
\begin{matrix}
\mathbb I & \mathbb O \\
\mathbb O & - \mathbb I 
\end{matrix}\right)$
there does not exist any small additive perturbation which can produce non-zero elements in the off-diagonal blocks of its square at linear level. 

The reason is simple. The unity matrix $\mathbb I$ does not single out any preferred directions. And one can arbitrarily introduce two subspaces with different signs of the eigenvalues for the square root. However, if we add a perturbation to $\mathbb I$ then, whatever small it is, it does produce preferred directions along its eigenvectors. And if the perturbation does not commute with our $\sqrt{\mathbb I}$ it means that the choice of two subspaces did not properly respect the geometry of the perturbation. And the mismatch can have arbitrarily large angles which prevent us from smoothly changing this particular square root.

Note however, that a smooth change of invariants is of course possible for all solutions. It is ensured by a whole manifold of such square roots which are connected by similarity transformations including rotations of the subspaces. Unfortunately, a simple inspection shows that perturbation theory around those solutions is more problematic than for the standard choice even in these terms \cite{paper2}. But it is probably the only hope to meaningfully deal with them at all.

To summarise, we propose to treat the action of massive gravity
\begin{equation}
\label{MGnew}
S=\int d^4 x \sqrt{-g} \left(R
+m^2 \sum_{n=0}^4 \beta_n {\mathfrak e}_n\right)
\end{equation}
such that the quantities ${\mathfrak e}_i$'s are not explicitly related to some square root matrices but rather defined as solutions of the following equations:
\begin{eqnarray}
\label{fraksta}
e_1(g^{-1}\eta) & = & {\mathfrak e}_1^2-2{\mathfrak e}_2,\\
e_2(g^{-1}\eta) & = & {\mathfrak e}_2^2-2{\mathfrak e}_1{\mathfrak e}_3+2{\mathfrak e}_4,\\
e_3(g^{-1}\eta) & = & {\mathfrak e}_3^2-2{\mathfrak e}_2{\mathfrak e}_4,\\
\label{frakend}
e_4(g^{-1}\eta) & = & {\mathfrak e}_4^2.
\end{eqnarray}
Note that it has nothing in common with another proposal to evade square roots in the action \cite{me} which made use of auxiliary fields $\Phi^{\mu}_{\nu}$ with a constraint that $\Phi^2=g^{-1}\eta$. The latter makes no good for the exotic square roots since at the end of the day the $\Phi$ field is nothing but the square root matrix with all its big problems.

We should note that there is also another way to avoid square roots in massive gravity, namely the vielbein formulation \cite{Kurt}. It has been shown equivalent to the metric approach as long as the "symmetric vielbein condition" is satisfied \cite{Cedric}. However, the latter is not strictly necessary, and therefore these are two different models. The difference is somewhat subtle, of course. But it appears to be important when it comes to discussions about generalised matter couplings \cite{Noller}. Leaving possible relations with vielbein formulations  for future work, this paper deals only with the metric version of massive gravity.

\section{Linearised massive gravity in the new method}

Let us now show how to use equations (\ref{fraksta}) -- (\ref{frakend}) to reproduce the Fierz-Pauli action. Note though again that for non-standard square roots the procedure would not go that simple \cite{paper2} but it is beyond the scope of the present paper.

Using the definitions (\ref{1L}), (\ref{2L}), (\ref{3L}), and possibly (\ref{4L}) we get for the left hand sides of our equations
\begin{eqnarray*}
e_1(g^{-1}\eta) & = & 4- h^{\mu}_{\mu}+ h_{\mu\nu}h^{\mu\nu}+{\mathcal O}(h^3),\\
e_2(g^{-1}\eta) & = & 6- 3 h^{\mu}_{\mu}+\frac12 (h^{\mu}_{\mu})^2+\frac52 h_{\mu\nu}h^{\mu\nu}+{\mathcal O}(h^3),\\
e_3(g^{-1}\eta) & = & 4- 3 h^{\mu}_{\mu}+ (h^{\mu}_{\mu})^2+2 h_{\mu\nu}h^{\mu\nu}+{\mathcal O}(h^3),\\
e_4(g^{-1}\eta) & = & 1- h^{\mu}_{\mu}+\frac12 (h^{\mu}_{\mu})^2+\frac12 h_{\mu\nu}h^{\mu\nu}+{\mathcal O}(h^3).
\end{eqnarray*}

We are interested in perturbations around the trivial solution of $\sqrt{\mathbb I}=\mathbb I$, and therefore we put
${\mathfrak e}_1=4+\delta{\mathfrak e}_1$, ${\mathfrak e}_2=6+\delta{\mathfrak e}_2$, ${\mathfrak e}_3=4+\delta{\mathfrak e}_3$, ${\mathfrak e}_4=1+\delta{\mathfrak e}_4$.

Obviously, equation (\ref{frakend}) for ${\mathfrak e}_4$ can straightforwardly be solved to any order we like. And the result
$${\mathfrak e}_4=1-\frac12 h^{\mu}_{\mu}+\frac18 (h^{\mu}_{\mu})^2+\frac14 h_{\mu\nu}h^{\mu\nu}+{\mathcal O}(h^3)$$
of course reproduces (\ref{e4}).

Then the other three equations give at the linear order:
\begin{eqnarray*}
h^{\mu}_{\mu} & = & 2\delta{\mathfrak e}_2-8\delta{\mathfrak e}_1,\\
2h^{\mu}_{\mu} & = & 8\delta{\mathfrak e}_1+8\delta{\mathfrak e}_3-12\delta{\mathfrak e}_2,\\
9h^{\mu}_{\mu} & = & 2\delta{\mathfrak e}_2-8\delta{\mathfrak e}_3
\end{eqnarray*}
which easily yields $\delta{\mathfrak e}_1=-\frac12 h^{\mu}_{\mu}$, $\delta{\mathfrak e}_2=-\frac32 h^{\mu}_{\mu}$, $\delta{\mathfrak e}_3=-\frac32 h^{\mu}_{\mu}$.

Substituting it back we get equations for the second order corrections
\begin{eqnarray*}
 h_{\mu\nu}h^{\mu\nu} - \frac14 (h^{\mu}_{\mu})^2 & = & 8\delta{\mathfrak e}^{(2)}_1-2\delta{\mathfrak e}^{(2)}_2,\\
2h_{\mu\nu}h^{\mu\nu} - \frac12 (h^{\mu}_{\mu})^2 & = & 12\delta{\mathfrak e}^{(2)}_2-8\delta{\mathfrak e}^{(2)}_1-8\delta{\mathfrak e}^{(2)}_3,\\
5h_{\mu\nu}h^{\mu\nu} + \frac74 (h^{\mu}_{\mu})^2 & = & 8\delta{\mathfrak e}^{(2)}_3-2\delta{\mathfrak e}^{(2)}_2
\end{eqnarray*}
and the solution is $\delta{\mathfrak e}^{(2)}_1=\frac38 h_{\mu\nu}h^{\mu\nu}$, $\delta{\mathfrak e}^{(2)}_2=h_{\mu\nu}h^{\mu\nu}+\frac18 (h^{\mu}_{\mu})^2$, $\delta{\mathfrak e}^{(2)}_3=\frac78 h_{\mu\nu}h^{\mu\nu}+\frac14 (h^{\mu}_{\mu})^2$. We see that the formulae (\ref{e1}) -- (\ref{e4}) are successfully reproduced.

Now, one only needs to plug it into the action (\ref{MGnew}), and the Fierz-Pauli theory is totally at hand.

\section{Conclusions}

Massive and bimetric gravity are a very active field of research. And actually, tremendous progress has been achieved in the recent years. However, some foundational issues remain poorly understood. And some of the most puzzling aspects touch the problem of square roots. In this paper we presented a new method of working with the model which deals directly with spectral invariants rather than with square root matrices themselves. As such, it might become beneficial for understanding the role of non-standard square roots which is the subject we hope to provide more details about very soon.

\vspace{4ex}
{\Large \textbf{Acknowledgments}}

\vspace{2ex}
AG enjoyed many inspiring discussions about the square roots and other topics of massive gravity with Fawad Hassan and Mikica Kocic. AG is grateful to the Dynasty Foundation for support; and also support of the Russian Foundation for Basic Research in the initial stages of the project under the grant 12-02-31214 is gratefully acknowledged.

\end{document}